# Improving Real-Time Bidding in Online Advertising Using Markov Decision Processes and Machine Learning Techniques

Parikshit Sharma[1]

[1]Department of Mathematics, Birla Institute of Technology and Science, Pilani, Rajasthan, India (333031)

[1]E-Mail: parikshitsharma2001@gmail.com

## ABSTRACT

Real-time bidding has emerged as an effective online advertising technique. With real-time bidding, advertisers can position ads per impression, enabling them to optimise ad campaigns by targeting specific audiences in real-time. This paper proposes a novel method for real-time bidding that combines deep learning and reinforcement learning techniques to enhance the efficiency and precision of the bidding process. In particular, the proposed method employs a deep neural network to predict auction details and market prices and a reinforcement learning algorithm to determine the optimal bid price. The model is trained using historical data from the iPinYou dataset and compared to cutting-edge real-time bidding algorithms. The outcomes demonstrate that the proposed method is preferable regarding cost-effectiveness and precision. In addition, the study investigates the influence of various model parameters on the performance of the proposed algorithm. It offers insights into the efficacy of the combined deep learning and reinforcement learning approach for real-time bidding. This study contributes to advancing techniques and offers a promising direction for future research.

## INTRODUCTION

**Real-time bidding (RTB)** lets marketers bid on ad impressions in real-time. A demand-side platform collects and analyses user data from display advertising websites to identify the best ad to display. Reinforcement learning-based real-time display advertising bidding was proposed by Cai et al. [1]. Their deep Q-network model determined the optimal bidding price in real-time auctions, outperforming existing methods. Karlsson [2] studied programmatic advertising feedback control approaches to optimise real-time bidding. Control theory was applied to proposal shading, budget allocation, and inventory management, with pros and cons. Liu and Yu [3] presented bid-aware active learning for display advertising real-time bidding. Their strategy improved bidding efficiency and precision using active learning and proposal prediction models. The authors tested their technique using real-world datasets.

**Display advertisements** sell products, services, and brands online through images, videos, and graphics. Banner adverts, pop-ups, and sponsored content are common on websites, social media platforms, and mobile apps. Thompkins [4] reviews a decade of online advertising research and highlights key conclusions and areas for further investigation. Choi et al. [5] analyse online display advertising industry literature, including economic and strategic consequences, and suggest future research. Nuara et al. [6] present probabilistic modelling to optimise multi-channel advertising campaigns due to interdependencies and unpredictability. Optimising bids and budgets improve display advertising campaigns. CTR, conversions, and ROI rise. Optimisation can help businesses enhance their advertising campaigns by discovering the best ad creatives, audiences, and channels. Geng et al. [7] suggested an automated bidding and budget optimisation system for performance advertising campaigns that use bidding and budget allocation algorithms to optimise campaign performance. Avadhanula et al. [8] proposed a stochastic bandit framework for optimising multi-platform advertising budgets. The method dynamically allocates budgets to platforms based on performance and estimation uncertainty. Luzon et al. [9] suggested optimising and learning budget allocation for social media advertising campaigns. Machine learning predicts the ideal budget allocation for the next period based on prior performance and budget allocation. Lin et al. [10] proposed a budget-constrained real-time bidding (BCRTB) optimisation algorithm that utilises multiple predictors to enhance performance.

**The iPinYou dataset** is a publicly accessible dataset containing a vast quantity of real-world data from the Chinese market. The dataset includes billions of ad impressions, user profiles, and contextual factors such as website type and time of day. Zhang et al. [11] evaluated the performance of various real-time bidding algorithms using the iPinYou benchmark dataset, which is extensively utilised in online advertising research. The iPinYou Global RTB Bidding Algorithm Competition Dataset, which contains four months of bidding logs from the iPinYou advertising platform, was created by Liao et al. [12] to facilitate research on RTB algorithms. Huang et al. [13] proposed a novel click-through rate prediction model based on deep and cross networks; its performance on the iPinYou dataset was promising.

**Deep Landscape Forecasting** can predict an advertisement's efficacy across various landscapes, including distinct user segments, ad positions, and contexts. RTB algorithms can make superior decisions regarding ad placement, bid price, and budget allocation when utilising DLF. Ren et al. [14] propose a deep landscape forecasting method for real-time bidding advertising to estimate the number of ad impressions that can be served to each user within the bidding landscape. Würfel et al. [15] present an interpretable deep learning method for online advertising revenue forecasting, which predicts future advertising revenues based on historical performance data. Ghosh et al. [16] present a scalable bid landscape forecasting method for real-time bidding that addresses the difficulty of predicting the landscape of the forthcoming auction in real-time.

**The Markov Decision Process (MDP)** is a mathematical framework for modelling decision-making problems with indeterminate action outcomes. It consists of states, actions that can be performed in each state, transition probabilities that specify the likelihood of transitioning from one state to another after acting, and rewards or costs associated with each transition. Du et al. [17] propose a constrained Markov decision process (CMDP) approach for enhancing real-time bidding, considering ad impression quality, budget constraints, and bidding strategy. Agrawal et al. [18] present a Markov decision model for administering display-advertising campaigns that employ a dynamic pricing mechanism to calculate optimal bid amounts. Shanahan and den Poel [19] propose a Markov decision process approach to determine online advertisements' optimal frequency limitation policy to maximise click-through rates. Boutilier and Lu [20] propose a budget allocation method that employs weakly coupled, constrained Markov decision processes to optimise resource allocation across multiple contending projects.

From the literature review done above, the following can be considered as the **novelty** of the present study:

- Most prior works viewed bid optimisation as a static optimisation problem involving either treating the value of each impression independently or setting a bid price for each segment of ad volume. However, the tendering for a particular ad campaign would occur repeatedly before the budget is exhausted. Consequently, proposal optimisation occurs in a real-time setting.
- Based on the novel modelling methodology, the present model can generate flexible forecasting auction results for each ad request without making any prior assumptions about the market price distribution, as demonstrated in the experiment.
- Using a comprehensive loss function for censorship management and going beyond traditional survival analysis methodologies to model the market price distribution more accurately.

## PROBLEM DEFINITION

We will use Markov decision process (MDP) to optimize the advertiser's total revenue. The problem is modelled as an auction where advertisers submit bids for ad impressions, and the ad exchange selects a winning bid to display an ad on a website. The goal is to maximize the advertiser's revenue while considering the budget constraints and the uncertain outcomes of user click behaviour.

The MDP is defined as a tuple $\langle S, A, R, P, \gamma \rangle$, where $S$ is the state space, $A$ is the action space, $R$ is the reward function, $P$ is the state transition probability function, and $\gamma$ is the discount factor. The state space is defined as $S = 0,1,2, \ldots, N$, where $N$ is the maximum number of impressions that can be shown to users, and the state $s_t$ at time $t$ represents the number of ad impressions that have been shown so far.

The action space $A$ is defined as the set of all possible bids that the advertiser can submit for an ad impression. The reward function $R$ is defined as $R(s_t, a_t) = c_t - a_t$, where $c_t$ is the expected revenue from a user clicking on the ad after seeing it, and $a_t$ is the bid submitted by the advertiser at time $t$. The transition probability function $P$ is defined as the probability that a user clicks on the ad after seeing it, which is estimated based on historical data.

The budget constraint is incorporated into the MDP by adding a penalty term to the reward function, which is defined as $-\infty$ if the sum of the bids submitted so far exceeds the advertiser's budget or if the ad is not clicked. The objective is to find a bidding policy that maximizes the expected total reward over a finite time horizon.

## PROBLEM FORMULATION

In the MDP formulation of RTB, let us consider a set of possible bids $B = b_1, b_2, \ldots, b_n$ that the agent can place for an ad impression. The agent's objective is to maximise its expected cumulative reward over a finite time horizon, given by:

$$\sum_{t=1}^{T} r_t$$

where $T$ is the time horizon and $r_t$ is the reward obtained at time step $t$. The reward is defined as:

$$\begin{cases} c_t - b_t, & \text{if the ad is clicked} \\ 0, & \text{otherwise} \end{cases}$$

where $c_t$ is the click-through rate of the ad impression at time step $t$, and $b_t$ is the bid placed by the agent. The click-through rate is modeled as a function of the features of the ad impression, denoted by $s_t$:

$$c_t = f(s_t)$$

The state of the environment is given by the features of the ad impression at time step $t$ denoted by $s_t$. The features can include information about the user, the website, the ad, and the context of the impression. The state evolves according to a probability distribution, denoted by $P(s_{t+1}|s_t, b_t)$, which depends on the bid placed by the agent.

The agent's action is the bid placed for the ad impression, denoted by $a_t$. The action space is discrete and consists of the set of possible bids $B$. The agent selects its action based on a policy $\pi(a_t|s_t)$, which is a probability distribution over the action space given the current state. The policy is learned by the agent using reinforcement learning.

Let us also introduce a budget constraint into the MDP formulation, which limits the total amount of money the agent can spend on bids over the time horizon. The budget constraint is given by:

$$\sum_{t=1}^{T} b_t \leq B_0$$

where $B_0$ is the initial budget of the agent. The budget constraint is incorporated into the MDP by adding a penalty term to the reward function:

$$r_t = \begin{cases} c_t - b_t, \text{ if the ad is clicked and } \sum_{i=1}^{t-1} b_i \leq B_0 \\ -\infty, \text{ if the ad is not clicked or } \sum_{i=1}^{t-1} b_i > B_0 \end{cases}$$

The penalty term ensures that the agent does not spend more than its budget over the time horizon.

## PROBLEM SOLUTION

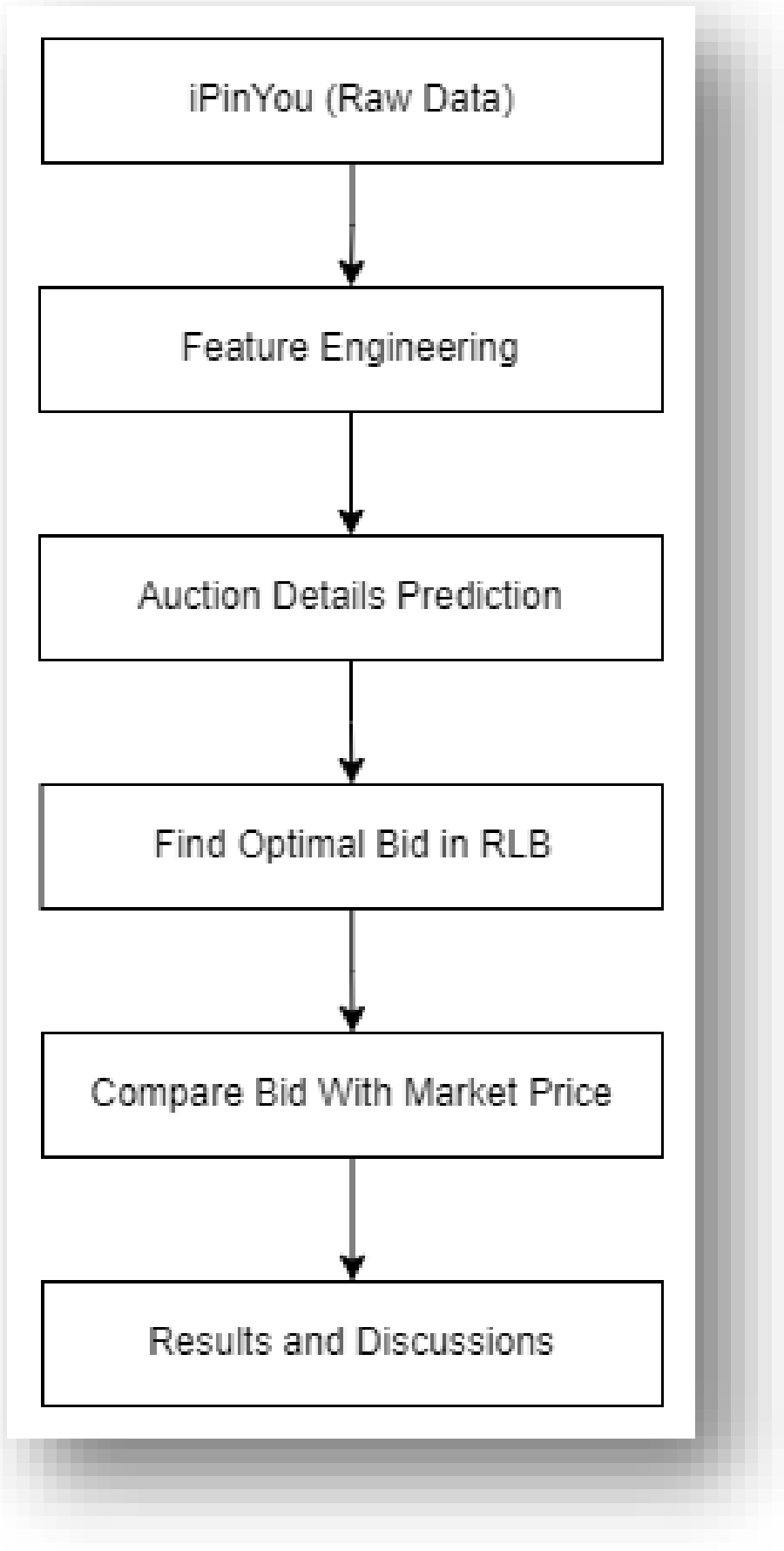

*Figure 1: Solution Methodology*

In this section, we delve into the details of the proposed Deep Landscape Forecasting (DLF) framework, which is a deep learning-based method that can forecast real-time bidding (RTB) advertising details using historical data. Figure 1 shows the solution methodology followed in this paper. The goal of landscape forecasting is to predict the future state of the ad auction landscape, which includes the bids placed by ad buyers and the characteristics of the ad slots available for auction. The DLF framework is composed of four main components: feature extraction, temporal modeling, landscape forecasting, and loss function. The input data to the DLF framework is a sequence of historical ad auction landscapes, denoted as $\{\boldsymbol{X_1}, \boldsymbol{X_2}, \ldots, \boldsymbol{X_T}\}$, where each $\boldsymbol{X_t}$ represents the ad auction landscape at time $t$.

**Feature Extraction**

We first extract low-level features from each ad auction landscape using convolutional neural networks (CNNs), denoted as $\{\boldsymbol{F_1}, \boldsymbol{F_2}, \ldots, \boldsymbol{F_T}\}$. The CNNs are trained to extract relevant information from the ad images, ad text, ad format, and ad placement. The output of each CNN is a feature map $\boldsymbol{F_t} \in R^{H \times W \times C}$, where $H$ and $W$ are the height and width of the feature map, respectively, and $C$ is the number of channels. Each channel corresponds to a specific feature of the ad auction landscape, such as the ad image quality, the ad text relevance, and the ad format popularity.

**Temporal Modeling**

Here we use a recurrent neural network (RNN) to model the temporal dependencies among the extracted features, denoted as $\{\boldsymbol{H_1}, \boldsymbol{H_2}, \ldots, \boldsymbol{H_T}\}$. The RNN takes the sequence of feature maps $\{\boldsymbol{F_1}, \boldsymbol{F_2}, \ldots, \boldsymbol{F_T}$ as input and produces a sequence of hidden states $\{\boldsymbol{H_1}, \boldsymbol{H_2}, \ldots, \boldsymbol{H_T}\}$. The RNN model is defined as follows:

$$\mathbf{h_t} = f(\mathbf{F_t}, \mathbf{h_{t-1}}),$$

where $\boldsymbol{h_t} \in R^D$ is the hidden state at time $t$, $D$ is the dimension of the hidden state, $f$ is a non-linear function such as the gated recurrent unit (GRU) or the long short-term memory (LSTM) unit, and $\boldsymbol{h_{t-1}}$ is the previous hidden state.

**Landscape Forecasting**

The main component of the DLF framework is the landscape forecasting module, which predicts the future state of the ad auction landscape based on the historical sequence of hidden states. The landscape forecasting module is implemented as a fully connected neural network, denoted as $g$, which takes the hidden state at each time step as input and produces a forecasted landscape $\widehat{\boldsymbol{X}_t}$:

$$\widehat{\mathbf{X}_t} = g(\mathbf{h_t}).$$

The landscape forecasting module can be trained using any standard supervised learning algorithm such as gradient descent or stochastic gradient descent. In this study, we have used stochastic gradient descent. The optimization of the objective function is performed using stochastic gradient descent with mini batches, which updates the parameters in each iteration as follows:

$$\theta_{t+1} = \theta_t - \eta_t \nabla_{\theta_t} L(\widehat{y_i}, y_i) - \eta_t \lambda \nabla_{\theta_t} \Omega(\theta_t)$$

where $\theta_t$ denotes the parameters at the $t^{th}$ iteration, $L(\widehat{y_i}, y_i)$ denotes the loss function used to compute the error between the predicted and true values, $\eta_t$ denotes the learning rate at the $t^{th}$ iteration, and $\nabla_{\theta_t}$ denotes the gradient with respect to the parameters.

**Loss Function**

We have used the mean squared error (MSE) loss as the objective function for training the DLF framework. The MSE loss measures the distance between the predicted ad auction landscape and the true ad auction landscape. Formally, the MSE loss is defined as follows:

$$\mathcal{L}(\{\mathbf{X_1}, \mathbf{X_2}, \dots, \mathbf{X_T}\}, \{\widehat{\mathbf{X}_1}, \widehat{\mathbf{X}_2}, \dots, \widehat{\mathbf{X}_T}\}) = \frac{1}{T}\sum_{t=1}^{T} |\mathbf{X_t} - \widehat{\mathbf{X}_t}|_2^2,$$

where $\boldsymbol{X_t}$ and $\widehat{\boldsymbol{X_t}}$ denote the actual and predicted values, respectively, at time step $t$. $|\cdot|_2$ is the $L2$ norm that measures the Euclidean distance between the actual and predicted values. The objective function is optimized using stochastic gradient descent with mini-batch training. In addition to the MSE loss, the authors also introduce a regularization term to prevent overfitting and improve generalization performance. The regularization term is defined as the sum of the Frobenius norm of the weights of the convolutional and fully connected neural networks. The total loss function is then defined as follows:

$$\mathcal{L}_{total} = \mathcal{L} + \lambda_1 \sum_{l \in conv} |\mathbf{W}_l|_F^2 + \lambda_2 \sum_{l \in fc} |\mathbf{W}_l|_F^2,$$

where $\lambda_1$ and $\lambda_2$ are regularization coefficients, $conv$ and $fc$ are the set of layers in the convolutional and fully connected neural networks, respectively, and $|\cdot|_F$ denotes the Frobenius norm.

**Bid Optimization**

Here we propose a dynamic programming approach for solving the optimal bidding problem in real-time bidding advertising. The problem is modelled as a Markov Decision Process (MDP) where at each auction, the advertiser selects a bid value that maximizes the expected total reward over a finite horizon. The state of the MDP is defined as the remaining budget and the number of remaining auctions, and the action is the bid value. The transition probabilities are defined as the probability of winning the auction given the bid value and the previous state. We use dynamic programming to solve this MDP by computing the optimal value function and the optimal policy. The value function is defined as the expected total reward starting from a given state and following the optimal policy. We will use a recursive formulation for the value function using the Bellman equation:

$$V^*(b,t) = \max_{x \in [0,b]} \{ r(x,b,t) + \sum_{s=0}^{\infty} \sum_{w \in \{0,1\}} p(w|x,b,t) P_{w,s} V^*(b-x, t-1+s) \}$$

where $V^*(b,t)$ is the optimal value function, $b$ is the remaining budget, $t$ is the remaining number of auctions, $x$ is the bid value, $r(x,b,t)$ is the immediate reward of bidding $x$ at time $t$ given the remaining budget $b$, $p(w \mid x,b,t)$ is the probability of winning the auction given the bid value $x$, the remaining budget $b$, and the remaining number of auctions $t$, $P_{w,s}$ is the probability of transitioning from time $t$ to time $t+1$ with $w$ indicating whether the advertiser wins the auction and $s$ indicating the number of auctions elapsed in between, and $V^*(b-x, t-1+s)$ is the optimal value function at time $t-1+s$ given that the advertiser wins the auction with probability $w$.

The optimal policy is then obtained by selecting the bid value that maximizes the value function at each state. The authors also propose a bid adjustment technique to ensure that the budget constraint is satisfied at each step. The dynamic programming approach provides an exact solution to the optimal bidding problem but is computationally expensive, especially when the number of auctions is large. To address this, a modified algorithm that uses a truncated value function and an approximate policy is used. The truncated value function only considers a finite number of future auctions, and the approximate policy selects the bid value that maximizes the truncated value function. This modified algorithm has a similar performance to the exact algorithm but with much lower computational complexity. Overall, the dynamic programming solution proposed in this section provides a rigorous and exact approach for solving the optimal bidding problem in real-time bidding advertising, and the modified algorithm offers a computationally efficient alternative.

## RESULTS AND DISCUSSIONS

### 1. Auction Market Price PDF Prediction

The auction details used to optimize the bidding price of an auction is predicted using Deep Landscape Forecasting (DLF). The probability density function of the market price is predicted using DLF and other six algorithms for comparison. The comparison id sone using Area Under the Curve (AUC), Logarithmic Loss (Log Loss), and Averaged Negative Log Probability (ANLP). AUC is the area under the Receiver Operating Characteristic (ROC) curve, which compares the True Positive Rate (TPR) to the False Positive Rate (FPR) at different classification thresholds. The AUC score ranges from 0 to 1, with greater values representing superior performance. Log Loss evaluates the performance of a model that produces probabilities by calculating the logarithm of the predicted probability of each class and summing these values for each observation in the test set. A lower Log Loss value indicates a more accurate model prediction. The average negative logarithm of the model's predicted probabilities for the correct class label is measured by ANLP. The model's performance is superior the lower the value of the metric. This metric is frequently applied to multi-class classification problems in which the model predicts the probabilities for each class and the class with the highest probability is selected as the predicted class. Table-1 shows the comparative analysis for the below discussed methods for 2259 campaign id.

**Kaplan-Meier (KM)** estimator is a non-parametric statistic used to estimate the probability of an event occurring over time, such as a click or a conversion in real-time bidding. In survival analysis, it is frequently used to estimate the survival function. In real-time bidding for display advertisements, the Kaplan-Meier estimator can be used to predict the market price by estimating the probability of winning an auction at various bid levels, taking into consideration competing bidders' maximum bids.

**Gamma distribution-based regression** model utilises ad position, advertiser, and user information to estimate the winning price's probability density function. The estimated density function is then fitted with a gamma distribution to predict the winning price. It represents the non-negative and asymmetrical characteristics of the auction bidding data. This model outperforms commonly employed regression models in terms of accuracy of prediction.

**Lasso-Cox** model is a form of survival analysis used to predict market prices in real-time bidding. It combines Cox's proportional hazards model, which predicts the time to an event, with Lasso regularisation, which identifies the most influential factors. The Adaptive Lasso algorithm is used to estimate the regression coefficients of the Cox model, allowing for automatic variable selection, and thereby enhancing the accuracy of prediction.

**DeepHit** models the joint distribution of time-to-event and event type and learns the mapping between input features and distribution parameters using neural networks. It has been demonstrated that the method outperforms traditional survival analysis models and has the potential to enhance real-time bidding strategy.

**Deep Weighted Poisson Regression (DWPP)** estimates the market price by combining a Poisson regression model with a neural network. The model addresses the issue of censoring using a weighted loss function and predicts the market price based on the input features. On multiple datasets, the method outperforms conventional regression models in terms of accuracy and efficiency.

**Recurrent neural network (RNN)** can capture the sequential character of bidding logs and can deal with sequences of variable length. The model uses historical bidding data to identify user behaviour patterns and predict the probability of obtaining a specific ad impression. This output is then fed back into the network for the subsequent time phase, enabling the network to make predictions based on the entire sequence. The RNN's internal memory enables it to recognise temporal dependencies in the data, making it an effective instrument for sequence modelling and prediction tasks.

| Algorithm | AUC | Log Loss | ANLP |
|---|---|---|---|
| **KM** | 0.680657 | 0.605585 | 14.81688 |
| **Gamma** | 0.514198 | 0.965122 | 7.939018 |
| **Cox** | 0.686271 | 0.887812 | 37.55281 |
| **DeepHit** | 0.689456 | 0.555907 | 5.764985 |
| **DWPP** | 0.702326 | 14.75364 | 41.49198 |
| **RNN** | 0.738268 | 0.791281 | 9.625792 |
| **DLF** | 0.838920 | 0.876278 | 5.244639 |

*Table 1: Comparative study of market price PDF prediction for 2259 campaign*

## 2. Campaign Bidding Price Optimization

The evaluation of the previously discussed algorithm for bid optimisation is conducted from the advertiser's campaign budget and lifespan (episode length) perspective. This algorithm is contrasted with five other algorithms discussed below. The comparative analysis outcomes are presented in Table-2. The parameters used for the model training are: $N = 1000, c_0 = 1/16$.

**Sponsored Search Markov Decision Process (SSMDP)** is a bidding strategy used for display advertising real-time bargaining. This approach represents the bidding problem as a Markov Decision Process, which can be solved using techniques from dynamic programming to determine the optimal bidding policy. The SSMDP method considers the click-through rate, conversion rate, and cost per click when determining the optimal bid to maximise the advertiser's anticipated revenue.

**Maximum Cost Per Click (MCPC)** is a common real-time bidding strategy for display advertisements. In this strategy, the advertiser sets a maximum bid limit for each ad impression, representing the most they are prepared to pay for a click. The winning offer is then determined based on the maximum CPC bid and the ad's click-through rate (CTR). This strategy enables advertisers to control their expenditures and ensures that they only pay for clicks on their advertisements up to their maximum CPC bid.

**Linear Bidding (LB)** is a frequently employed method for display advertisements. It entails estimating the value of a user impression based on the characteristics of the advertisement and the user, and then bidding a proportional amount. The optimal proposal is determined by balancing the anticipated revenue from the advertisement with the cost of its display. This strategy is easy to implement and can be effective in many situations, but it may not be optimal in more complex circumstances.

**Constrained Markov Decision Process (CMDP)** is a bidding strategy utilised in display advertising real-time bargaining. It is an extension of the standard MDP that permits the incorporation of state space constraints. The CMDP framework is intended to allow advertisers to maximise profits while adhering to budgetary and other constraints. This strategy is especially useful for advertisers with limited budgets, as it allows them to control their spending while optimising their bids. By incorporating constraints into the MDP framework, the CMDP approach provides advertisers seeking to maximise return on investment in real-time bidding for display advertising with a powerful and adaptable instrument.

**Batch Constrained Markov Decision Process (BCMDP)** is a bidding strategy used in real-time bidding for display advertisements to determine the optimal proposal. It is based on the Constrained Markov Decision Process (CMDP) and considers advertisers' budget constraints. In BCMDP, a group of auctions are evaluated concurrently, and a set of constraints are formulated for the group. Solving the CMDP problem under the provided constraints determines the optimal bid for each auction. The resulting bids are then distributed based on a predetermined allocation formula, such as the VCG mechanism. It has been demonstrated that the BCMDP strategy can significantly increase advertisers' expected revenue while adhering to budgetary constraints.

| Campaign | Algorithm | Objective | Auction | Impression | Clicks | Cost | Win Rate | CPM | E-CPC |
|---|---|---|---|---|---|---|---|---|---|
| 1458 | SSMDP | 20 | 350000 | 82043 | 20 | 1500065 | 23.44% | 18.28 | 75.00 |
| | MCPC | 218 | 350000 | 51853 | 218 | 1388007 | 14.82% | 26.77 | 6.37 |
| | LB | 255 | 350000 | 48009 | 255 | 1160457 | 13.72% | 24.17 | 4.55 |
| | RLB | 272 | 350000 | 45429 | 272 | 1406711 | 12.98% | 30.97 | 5.17 |
| | CMDP | 466 | 350000 | 89162 | 466 | 2536626 | 14.51% | 28.45 | 5.44 |
| | BCMDP | 462 | 350000 | 108651 | 462 | 2515175 | 17.68% | 23.15 | 5.44 |
| 2259 | SSMDP | 11 | 350000 | 92391 | 11 | 2024398 | 26.40% | 21.91 | 184.04 |
| | MCPC | 9 | 350000 | 38003 | 9 | 2035388 | 10.86% | 53.56 | 226.15 |
| | LB | 14 | 350000 | 46743 | 14 | 2035330 | 13.36% | 43.54 | 145.38 |
| | RLB | 18 | 350000 | 67731 | 18 | 2016836 | 19.35% | 29.78 | 112.05 |
| | CMDP | 22 | 350000 | 83757 | 22 | 2378214 | 20.08% | 28.39 | 108.10 |
| | BCMDP | 19 | 350000 | 81593 | 19 | 2302661 | 19.56% | 28.22 | 121.19 |
| 2261 | SSMDP | 16 | 343862 | 121913 | 16 | 1921384 | 35.45% | 15.76 | 120.09 |
| | MCPC | 12 | 343862 | 46272 | 12 | 1926368 | 13.46% | 41.63 | 160.53 |
| | LB | 17 | 343862 | 83772 | 17 | 1739679 | 24.36% | 20.77 | 102.33 |
| | RLB | 17 | 343862 | 88878 | 17 | 1923796 | 25.85% | 21.65 | 113.16 |
| | CMDP | 15 | 343862 | 83284 | 15 | 1925641 | 24.22% | 23.12 | 128.38 |
| | BCMDP | 17 | 343862 | 84764 | 17 | 1925641 | 24.65% | 22.72 | 113.27 |
| **2821** | SSMDP | 30 | 350000 | 91672 | 30 | 1937519 | 26.19% | 21.14 | 64.58 |
| | MCPC | 23 | 350000 | 41090 | 23 | 1952973 | 11.74% | 47.53 | 84.91 |
| | LB | 31 | 350000 | 56528 | 31 | 1495739 | 16.15% | 26.46 | 48.25 |
| | RLB | 36 | 350000 | 71847 | 36 | 1937604 | 20.53% | 26.97 | 53.82 |
| | CMDP | 65 | 350000 | 145906 | 65 | 3693887 | 22.04% | 25.32 | 56.83 |
| | BCMDP | 55 | 350000 | 116136 | 55 | 3693887 | 17.54% | 31.81 | 67.16 |
| **2997** | SSMDP | 115 | 156063 | 58866 | 115 | 613597 | 37.72% | 10.42 | 5.34 |
| | MCPC | 82 | 156063 | 29034 | 82 | 614884 | 18.60% | 21.18 | 7.50 |
| | LB | 77 | 156063 | 38978 | 77 | 270386 | 24.98% | 6.94 | 3.51 |
| | RLB | 119 | 156063 | 57267 | 119 | 609392 | 36.69% | 10.64 | 5.12 |
| | CMDP | 97 | 156063 | 49298 | 97 | 411088 | 31.59% | 8.34 | 4.24 |
| | BCMDP | 102 | 156063 | 50266 | 102 | 428397 | 32.21% | 8.52 | 4.20 |
| **3358** | SSMDP | 15 | 300928 | 52674 | 15 | 1722296 | 17.50% | 32.70 | 114.82 |
| | MCPC | 144 | 300928 | 23633 | 144 | 1699557 | 7.85% | 71.91 | 11.80 |
| | LB | 213 | 300928 | 16061 | 213 | 999776 | 5.34% | 62.25 | 4.69 |
| | RLB | 219 | 300928 | 24182 | 219 | 1683775 | 8.04% | 69.63 | 7.69 |
| | CMDP | 223 | 300928 | 53850 | 223 | 1438932 | 6.73% | 26.72 | 6.45 |
| | BCMDP | 201 | 300928 | 54361 | 201 | 1439570 | 7.62% | 26.48 | 7.16 |
| **3386** | SSMDP | 20 | 350000 | 80085 | 20 | 1673629 | 22.88% | 20.90 | 83.68 |
| | MCPC | 35 | 350000 | 34657 | 35 | 1681835 | 9.90% | 48.53 | 48.05 |
| | LB | 56 | 350000 | 41952 | 56 | 1555576 | 11.99% | 37.08 | 27.78 |
| | RLB | 65 | 350000 | 46871 | 65 | 1670203 | 13.39% | 35.63 | 25.70 |
| | CMDP | 72 | 350000 | 79660 | 72 | 1574648 | 12.04% | 19.77 | 21.87 |
| | BCMDP | 64 | 350000 | 81302 | 64 | 1571482 | 12.69% | 19.33 | 24.55 |
| **3427** | SSMDP | 11 | 350000 | 79358 | 11 | 1763480 | 22.67% | 22.22 | 160.32 |
| | MCPC | 91 | 350000 | 37213 | 91 | 1761734 | 10.63% | 47.34 | 19.36 |
| | LB | 170 | 350000 | 31432 | 170 | 1378162 | 8.98% | 43.85 | 8.11 |
| | RLB | 191 | 350000 | 37648 | 191 | 1711180 | 10.76% | 45.45 | 8.96 |
| | CMDP | 203 | 350000 | 80480 | 203 | 1759063 | 9.04% | 21.86 | 8.67 |
| | BCMDP | 195 | 350000 | 82125 | 195 | 1752960 | 11.72% | 21.35 | 8.99 |
| **3476** | SSMDP | 24 | 350000 | 77240 | 24 | 1724007 | 22.07% | 22.32 | 71.83 |
| | MCPC | 67 | 350000 | 30776 | 67 | 1731727 | 8.79% | 56.27 | 25.85 |
| | LB | 146 | 350000 | 20877 | 146 | 1039774 | 5.96% | 49.80 | 7.12 |
| | RLB | 139 | 350000 | 33466 | 139 | 1710701 | 9.56% | 51.12 | 12.31 |
| | CMDP | 136 | 350000 | 67558 | 136 | 1963504 | 7.93% | 29.06 | 14.44 |
| | BCMDP | 143 | 350000 | 78790 | 143 | 1963257 | 7.15% | 24.92 | 13.73 |

*Table 2: Comparative analysis for bid optimization*

## CONCLUSION

The combination of deep learning and reinforcement learning has shown great promise for enhancing the efficacy and precision of real-time bidding in display advertising. Deep neural networks, specifically recurrent neural networks, enable accurate modelling and forecasting of market prices, whereas reinforcement learning enables adaptive and dynamic pricing strategies in response to changing market conditions. Moreover, the incorporation of contextual information, such as user behaviour and ad attributes, has proven effective for enhancing bidding performance. Nevertheless, there are still obstacles to surmount, such as the complexity and computational costs of deep learning models and the need for more advanced contextual information extraction and processing techniques. Notwithstanding, these research efforts have paved the way for further investigation and development of real-time advertising systems that can more effectively leverage the power of deep learning and reinforcement learning to optimise ad placements and increase revenue for advertisers and publishers.